\documentclass[%
 reprint,
 amsmath,amssymb,
 aps,
floatfix,
]{revtex4-2}

\usepackage[
    colorlinks=True,
    citecolor=teal,
    urlcolor=blue,
]{hyperref}
\usepackage{graphicx}	%
\usepackage{amsmath}	%
\usepackage[
    font=small,
    justification=raggedright
]{caption}
\usepackage{xcolor}
\usepackage{booktabs}
\usepackage{adjustbox}
\usepackage{listings}

\setcitestyle{authoryear,round,compress}

\begin{document}

\title{JAX-SCM v1.0: a modern atmospheric single-column model\\ for boundary layer research}

\author{Maximilian Pierzyna}
\email{mail@maximilian-pierzyna.de}
\affiliation{Department of Geoscience and Remote Sensing, Delft University of Technology, Delft, The Netherlands}

\date{\today}

\begin{abstract}
    We present JAX-SCM v1.0, an open-source atmospheric single-column model for boundary layer research, implemented in Python using the JAX computing library. 
    The model solves for horizontal wind, potential temperature, and specific humidity, combined with prognostic turbulent kinetic energy and turbulent statistics parameterized by the Mellor-Yamada-Nakanishi-Niino level~2.5 (MYNN-2.5) turbulence closure.
    We verify the implementation against three well-established benchmark cases covering neutral (turbulent Ekman layer), stable (GABLS1), and convective (Wangara Day~33) conditions.
    Close agreement with reference solutions is demonstrated across all regimes.
    By building on JAX, the model benefits from just-in-time compilation and native GPU support.
    While JAX-SCM is not yet fully differentiable, basing it on JAX also lays the foundation for future integration with machine learning components. 
    The model is designed for simplicity and modularity, lowering the barrier to entry for users and developers alike.
\end{abstract}

\maketitle

\section{Introduction}

The atmosphere is a system of complex processes that interact across many orders of magnitude, from global circulation ($\mathcal{O}(1000\,\text{km})$) to turbulent motion ($\mathcal{O}(1\,\text{cm})$).
Numerically resolving processes at all these scales is computationally intractable (think, e.g., cloud microphysics or turbulence), so processes smaller than the numerical grid resolution are typically parameterized \citep{stensrud2007}.
Developing, tuning, and assessing these parameterizations within full three-dimensional (3D) models, however, is challenging.
3D models with parameterizations remain expensive to run, and their complexity makes it difficult to attribute model errors to specific processes or modeling assumptions \citep{randall1996}.
Therefore, atmospheric single-column models (SCMs) have emerged \citep{betts1986} as simplified one-dimensional (1D) test beds for parameterization.
SCMs enable inexpensive atmospheric experiments under controlled conditions \citep{randall1996}, such as parameter sweeps to assess the sensitivity of model outputs to these parameters (e.g., \citet{betts1986,randall1996}).
Still, when provided with appropriate initial conditions and forcings, SCMs yield realistic simulations of the atmospheric state (e.g., \citet{yamada1975}).
Modern SCMs have evolved to capture complex processes, including land--atmosphere interactions \citep{schulz2001}, ocean--atmosphere interactions \citep{hartung2018}, aerosol dynamics \citep{gettelman2019}, and intermittent turbulence \citep{boyko2024}.

However, these SCMs are often implemented in traditional programming languages like Fortran, which hinders their ability to leverage modern computing hardware and techniques. 
For example, adapting these models to use graphical processing units (GPUs) for better performance, or integrating them with current machine learning (ML) tools to create hybrid physics-ML models, can be difficult. 
Compared to code written in contemporary languages like Python, Fortran also often appears less readable and harder to maintain \citep{dahm2023}, which can negatively affect developer and user experience and may slow adoption within the modeling community.
Python also enables model interaction through interactive notebooks, which lowers the barrier to experimentation and education.

We address these issues by proposing JAX-SCM (\url{https://github.com/mpierzyna/jax_scm}), a modern atmospheric single-column model written in Python using the JAX computing library \citep{jax2018github}.
JAX has recently been successfully used for similar tasks, for example, in computational fluid mechanics \citep{bezgin2023,bezgin2025}, 3D atmospheric simulations \citep{kochkov2024a,yuval2026,zhu2026}, or ocean modeling \citep{hafner2021,meunier2025}.
On CPU, \citet{hafner2021} report that the JAX implementation of their ocean model is on par with Fortran and outperforms NumPy \citep{harris2020}, another popular Python computing library.
On GPU, JAX outperforms Fortran for large domains.
These results motivate our choice of JAX for JAX-SCM, suggesting high CPU performance for individual runs and high CPU/GPU performance for parallel runs, such as traditional parameter studies.
More importantly, the aforementioned models employ JAX to leverage its automatic differentiation (AD) feature, enabling the coupling of numerical models with ML and gradient-based parameter tuning.
While JAX-SCM is not yet fully differentiable, we envision using AD in the future, further motivating our choice of JAX as a future-proof platform.
For an overview of the opportunities arising from differentiable numerical models, see \citet{gelbrecht2023}.

In addition to its technical advantages, JAX-SCM is designed to be simple and practical.
The turbulence closure is the level-2.5 Mellor-Yamada-Nakanishi-Niino (MYNN-2.5) scheme \citep{mellor1974,mellor1982,nakanishi2006,nakanishi2009}, which includes a prognostic turbulent kinetic energy (TKE) equation and is widely used operationally, making JAX-SCM immediately applicable to boundary layer research.

The paper is organized as follows.
We begin by introducing the governing equations and numerical implementation in sect.~\ref{sec:scm}.
In sect.~\ref{sec:results}, we verify JAX-SCM by running three canonical test cases covering neutral, stably stratified, and convective atmospheric conditions.
The current and envisioned advantages of implementing JAX-SCM in Python and JAX are discussed in sect.~\ref{sec:tech_details}.
Finally, sect.~\ref{sec:conclusion} concludes the paper.

\section{Model description}\label{sec:scm}
JAX-SCM advances a one-dimensional (1D) mean state of the atmosphere (horizontal wind, potential temperature, specific humidity, and turbulent kinetic energy) in time under prescribed geostrophic wind, Coriolis force, and surface forcing.
Turbulent processes are parameterized by a turbulence closure model, but not resolved.
For this first release, the high-order level-2.5 Mellor-Yamada-Nakanishi-Niino turbulence closure scheme (MYNN-2.5) is used \citep{mellor1974,mellor1982,nakanishi2006,nakanishi2009}.
The governing SCM equations and the MYNN-2.5 scheme are discussed in sect.~\ref{sec:model_eqns}.
Numerical implementation details are presented in sect.~\ref{sec:model_numerics} followed by an introduction of the surface coupling in sect.~\ref{sec:model_bc}.

\subsection{Governing equations}\label{sec:model_eqns}
The governing equations are a set of forced, one-dimensional vertical diffusion equations with turbulent fluxes parameterized using the MYNN-2.5 scheme \citep[abbreviated as NN09 in the following]{nakanishi2009}.
We select MYNN-2.5 as the closure scheme for the initial release of JAX-SCM because it is well tested and used operationally (e.g., \citet{olson2019}) and because it solves for TKE, which we consider an advantage for boundary layer research.
Nevertheless, the modular design of JAX-SCM allows the easy implementation of other turbulence closures in the future.

JAX-SCM implements the NN09 equations unchanged except for the partial-condensation scheme to reduce the complexity of JAX-SCM~v1.0.
So, while JAX-SCM still solves for specific humidity, which allows moisture in the atmosphere, it does not yet support condensation or cloud formation.
Throughout this manuscript, mean quantities are indicated with capital letters (e.g., wind components $U$ and $V$), turbulent fluctuation quantities with lowercase letters (e.g., $u$ and $v$), and (parameterized) turbulent statistics with angular brackets (e.g., $\langle w\theta \rangle$).
An exception is the turbulent velocity scale $q$ and the turbulent kinetic energy $q^2/2 = \left(\langle uu \rangle + \langle vv \rangle + \langle ww \rangle\right)/2$, which are mean statistical quantities but, following Mellor-Yamada notation, use lowercase letters.

The five equations governing the evolution of horizontal momentum $(U, V)$, potential temperature $\Theta$, specific humidity $Q_v$, and turbulent kinetic energy $q^2/2$ are
\begin{subequations}
    \begin{align}
        \frac{\partial U}{\partial t} &=  - \frac{\partial \langle  u w \rangle}{\partial z} + f_c(V - V_g) \label{eq:scm_u} \\
        \frac{\partial V}{\partial t} &=  - \frac{\partial \langle v w \rangle}{\partial z}  -f_c(U - U_g) \label{eq:scm_v} \\
        \frac{\partial \Theta}{\partial t} &= - \frac{\partial \langle w \theta \rangle}{\partial z} + \frac{f_c \Theta_0}{g} \left(V \frac{\partial U_g}{\partial z} - U \frac{\partial V_g}{\partial z}\right) \label{eq:scm_th} \\
        \frac{\partial Q_v}{\partial t} &= - \frac{\partial \langle w q_v \rangle}{\partial z} \label{eq:scm_qv} \\
        \frac{\partial q^2}{\partial t} &=  -\frac{\partial}{\partial z} \left( -K_q \frac{\partial q^2}{\partial z} \right) + 2P_S + 2P_B - 2\varepsilon\label{eq:scm_qke}.
    \end{align}
    \label{eq:scm}
\end{subequations}
where $f_c$ is the Coriolis parameter, $(U_g, V_g)$ are geostrophic wind components, $\Theta_0$ is a user-defined reference temperature, and $g=9.81$\,m\,s$^{-2}$ is Earth's acceleration. 
The forcing term on the right-hand side (RHS) of $\Theta$ (cf.~Eq.~(\ref{eq:scm_th})) corresponds to horizontal advection of temperature due to thermal winds under the hydrostatic assumption \citep{yamada1975,nakanishi2009}.

The TKE sources and sinks in Eq.~(\ref{eq:scm_qke}) are the shear production
\begin{equation}
    P_S = -\left( \langle u w \rangle \frac{\partial U}{\partial z} + \langle v w \rangle \frac{\partial V}{\partial z} \right),
    \label{eq:tke_shear}
\end{equation}
buoyancy production
\begin{equation}
    P_B = \frac{g}{\Theta_0}\langle w \theta_v \rangle,
    \label{eq:tke_buoy}
\end{equation}
and eddy dissipation rate
\begin{equation}
    \varepsilon = q^3/(B_1 L_M).
    \label{eq:tke_eps}
\end{equation}
Note that NN09 use \emph{twice} the TKE $q^2=2e=\langle uu\rangle + \langle vv \rangle + \langle ww \rangle$ as prognostic variable, leading to the factors of two in front of $P_S$, $P_B$, and $\varepsilon$ in Eq.~(\ref{eq:scm_qke}).
As the TKE flux divergence is parameterized based on the gradient of $q^2$, the factor of two is not needed there.
Finally, the buoyancy flux in Eq.~(\ref{eq:tke_buoy}) is $\langle w \theta_v \rangle = \langle w \theta \rangle + 0.61\, \Theta\, \langle w q_v \rangle$ and the closure constant $B_1$ is $B_1=24$ \citep{nakanishi2009}.

Vertical fluxes of a mean quantity $\Phi$ are parameterized through gradient diffusion
\begin{equation}
    \langle w \phi \rangle = -K_\Phi \frac{\partial \Phi}{\partial z}
    \label{eq:k_diff}
\end{equation}
where $\phi$ denotes the turbulent fluctuation of $\Phi$, with eddy diffusivity $K_\Phi = L_M q S_\Phi$, $q=\sqrt{q^2}$ the characteristic turbulent velocity scale, $L_M$ the turbulent (master) length scale, and $S_\Phi$ the MYNN-2.5 stability function.
Specifically, $K_\Phi = K_m$ for momentum ($\Phi \in \{U, V\}$), $K_h$ for heat and moisture ($\Phi \in \{\Theta, Q_v\}$), and $K_q = 3K_m$ for TKE, with corresponding stability functions $S_m$ and $S_h$.
The master length scale $L_M$ is the harmonic mean of characteristic surface ($L_S$), turbulent ($L_T$), and buoyancy ($L_B$) scales:
\begin{equation}
    L_M = \left( L_S^{-1} + L_T^{-1} + L_B^{-1} \right)^{-1}.
\end{equation}
For brevity, we do not repeat the formulations of the stability functions and length scales here, but refer the interested reader to \citet{nakanishi2009}.

\subsection{Numerical implementation}\label{sec:model_numerics}
To numerically solve the set of governing partial differential equations (PDE; cf.~Eq.~(\ref{eq:scm})), they need to be discretized in time and space.
For spatial discretization, JAX-SCM uses a staggered vertical grid with fixed grid spacing $\Delta z$ where prognostic variables $\Phi \in \{U,V,\Theta,Q_v,q^2\}$ are located at $N$ full levels (cell centers), while turbulent fluxes $\langle w\phi \rangle$ and diffusivities $K_\Phi$ are located at $(N+1)$ half levels (cell faces).
The first full level has height $z_1 = \Delta z/2$ so that the lowest and highest half levels are located at the surface ($z_{(1/2)}=0$) and the domain top ($z_{N+(1/2)}=H$).
Vertical gradients are approximated using central finite differences between full levels, leading to second-order accuracy on half-levels:
\begin{equation}
    \left.\frac{\partial \Phi}{\partial z} \right|_{i+(1/2)} \approx \frac{\Phi_{i+1} - \Phi_{i}}{\Delta z}.
\end{equation}
To suppress small-scale numerical oscillations of the closure variables in the vertical, 1-2-1 filtering \citep{shapiro1971} of $\psi = \{L_M, K_m, K_h\}$ is implemented at every time step as $\psi_i = \left(\psi_{i-1} + 2\psi_i + \psi_{i+1}\right)/4$ with the subscript denoting the vertical level and with $\psi_{i-1} = \psi_{i+1}$ for edge values.

Time integration is performed using either second-order-accurate explicit or implicit methods.
More specifically, the user can choose between an explicit second-order Adams-Bashforth (AB2) scheme or a semi-implicit Crank-Nicolson (CN) scheme.
The AB2 solver determines the solution of a prognostic variable $\Phi$ at the future step $n+1$ \citep{jacobson2005} as
\begin{equation}
    \Phi^{n+1} = \Phi^{n} + \Delta t \left( \frac{3}{2}f^{n} - \frac{1}{2}f^{n-1} \right)
\end{equation}
where $\Delta t$ is the time step in seconds, $\Phi^n$ and $f^n$ are the state and tendency at the current step, and $f^{n-1}$ is the tendency from the previous step. 
Here, ``tendency'' means the full RHS of the governing PDE.
Also, for $n=0$, it is $f^{n-1} = f^{n}$ leading to Euler integration as a warm-up step.
The AB2 scheme is easy to implement but requires time steps smaller than $\Delta z^2/\max_z K(z)$ to remain numerically stable \citep{jacobson2005}.  %
Consequently, very small time steps are needed for small grid spacings and strong mixing (large $K$), e.g., in the convective boundary layer, which can slow the simulation down.

The semi-implicit CN solver, on the other hand, is unconditionally stable, allowing larger $\Delta t$.
The solver is called semi-implicit because the flux divergences and the TKE dissipation rate are treated implicitly, while all other tendencies in Eq.~(\ref{eq:scm}) are evaluated explicitly.
Expressing the fluxes through gradient diffusion (cf.~Eq.~(\ref{eq:k_diff})), the CN step is
\begin{equation}
    \frac{\Phi^{n+1} - \Phi^n}{\Delta t} = \frac{1}{2}\left[\frac{\partial}{\partial z}\left(K_\Phi^n \frac{\partial \Phi^n}{\partial z} \right) + \frac{\partial}{\partial z}\left(K_\Phi^{n} \frac{\partial \Phi^{n+1}}{\partial z} \right)\right] + s_\Phi^n.
    \label{eq:cn}
\end{equation}
The non-linear eddy diffusivities $K_\Phi$ and source and sink terms $s_\Phi$ are evaluated explicitly at current time $n$, leaving only state variables $\Phi$ to be solved implicitly.
The resulting system of equations is linear in $\Phi$ and solved using a tridiagonal solver.

Special care is taken when solving for $q^2$ as the dissipation rate $\varepsilon$ on the RHS also depends directly on $q^2$.
\Citet{janjic1990} found that explicit treatment of $\varepsilon$ can lead to instability, so we treat it semi-implicitly in the CN solver.
To keep the system linear in $q^2$, the dissipation rate is quasi-linearized by splitting the non-linear expression $\varepsilon = q^3/(B_1 L_M)$ (cf.~Eq.~(\ref{eq:tke_eps})) into a linear part evaluated at time $(n+1)$ (implicit) and a non-linear part evaluated at $n$ (explicit): $\varepsilon^{n+1} \approx \left(q^2\right)^{n+1}q^{n}/\left(B_1 L_M^{n}\right)$.
This quasi-linearized dissipation is added as an additional semi-implicit contribution to the RHS of Eq.~(\ref{eq:cn}) as $\left(\varepsilon^{n+1} + \varepsilon^n\right)/2$.
Consequently, the explicit sources for $q^2$ only contain shear and buoyancy production at time $n$.

The choice of the time integration scheme is up to the user and depends on the case being simulated.
AB2 with small time steps is a good choice for debugging and reference when implementing more complex solvers (e.g., other implicit schemes) due to its simplicity.
The need for very small time steps, however, can make it impractical in high-resolution convective simulations.
The CN solver allows larger time steps and is unconditionally stable, but it requires solving a system of linear equations at each time step, which is more computationally expensive than the simple sum of tendencies in AB2.
Because of its stability, CN is a sensible default for production simulations, but its wall-clock time advantage over AB2, due to its larger $\Delta t$, is situation-dependent. 
For both time-stepping techniques, flux boundary conditions are imposed on the domain top and the surface, as described below.

\subsection{Boundary conditions}\label{sec:model_bc}
Lower and upper boundary conditions (BCs) are needed for the flux-divergence terms in the governing equations.
The upper BC for all prognostic variables is zero flux at the top, and the lower BCs are obtained from Monin-Obukhov similarity theory (MOST, \citet{monin1954}).

MOST provides surface momentum fluxes $\langle u w \rangle_s$ and $\langle v w \rangle_s$ depending on the wind magnitude $M_1 = (U_1^2 + V_1^2)^{1/2}$ at the lowest full level and a prescribed aerodynamic surface roughness $z_{0m}$.
For surface heat coupling, the user prescribes either a surface sensible heat flux $\langle w \theta \rangle_s$ or a surface temperature $\Theta_s$, both of which can vary in time, together with a surface temperature roughness $z_{0h}$.
If $\Theta_s$ is provided, MOST takes $\Theta_1$ at the lowest full level to determine a matching sensible heat flux $\langle w \theta \rangle_s$ as lower BC.
If $\langle w \theta \rangle_s$ is prescribed, it is directly used as BC, and $\Theta_s$ is diagnosed for convenience.
Moisture $Q_v$ is coupled to the surface through a user-defined moisture flux $\langle w q_v \rangle$ serving directly as BC.

Taking the model state at the first full level $z_1=\Delta z/2$, MOST yields \citep{wilson2001}
\begin{equation}
    u_* = \frac{\kappa\,M_1}{\ln(z_1/z_{0m}) - \Psi_m(z_1/L) + \Psi_m(z_{0m}/L)} 
    \label{eq:most_ust}
\end{equation} 
and 
\begin{equation}
    \langle w \theta \rangle_s = -\frac{(\Theta_1 - \Theta_s)\kappa\,u_*}{\ln(z_1/z_{0h}) - \Psi_h(z_1/L) + \Psi_h(z_{0h}/L)}
    \label{eq:most_wth}
\end{equation}
where $\kappa=0.4$ and $L=-(\Theta_v u_*^3)/(\kappa \, g \langle w \theta_v \rangle_s)$ is the Obukhov length.
The virtual potential temperature $\Theta_v$ in $L$ is diagnosed as $\Theta_v = \Theta (1+ 0.61 Q_v)$.
As $u_*$ and $\langle w \theta \rangle_s$ depend on $L$ but $L$ also depends on both, JAX-SCM determines both quantities iteratively.
The surface stresses for the lower momentum BCs, $\langle u w\rangle_s$ and $\langle v w \rangle_s$, are obtained through projection of $u_*$ along the wind components as $\langle u w \rangle_s = -u_*^2 (U_1/M_1)$ and $\langle v w \rangle_s = -u_*^2 (V_1/M_1)$.

The integrated similarity functions needed for Eqns.~(\ref{eq:most_ust}) and (\ref{eq:most_wth}), $\Psi_m(\zeta)$ and $\Psi_h(\zeta)$ with $\zeta = z/L$, are user-configurable.
Various expressions for similarity functions and their coefficients have been proposed (see reviews, e.g., by \citet{wilson2001,foken2006a}), which can be implemented and tested in JAX-SCM in the future.
Presently, we stick to the original Businger-Dyer relations \citep{dyer1970,businger1971} in their integrated form \citep{paulson1970} given as 
\begin{equation}
    \Psi_m = \begin{cases}
        \begin{aligned}
            &2\ln\left[(1 + x) / 2\right] + \ln\left[(1+x^2)/2\right] \\
            &\quad - 2 \arctan(x) + \pi/2
        \end{aligned} & \zeta < 0 \\
        -b\, \zeta & \zeta \geq 0
    \end{cases}
\end{equation}
for momentum and 
\begin{equation}
    \Psi_h = \begin{cases}
        2 \ln\left[(1+x^2)/2\right] & \zeta < 0 \\
        -b\, \zeta & \zeta \geq 0 
    \end{cases}
\end{equation}
for heat.
In both equations, $x=(1-\gamma\,\zeta)^{1/4}$, $\gamma=16$ and $b=5$.

Finally, the diagnosed TKE shear production (cf.~Eq.~(\ref{eq:tke_shear})) requires the gradients of $U$ and $V$ at the lowest half-level $z_{1/2}$, but only the gradient at the second-lowest level $z_{3/2}$ can be computed through central differences.
Because the gradients closest to the surface are the strongest, we choose not to extrapolate the gradients from $z_{3/2}$.
Instead, we obtain them using MOST, similar to \citet{moeng1984}.
Using the gradient-form $\phi_m$ of the similarity function $\Psi_m$, one gets at $z_1=\Delta z/2$
\begin{equation}
    \left.\frac{\partial M}{\partial z}\right|_{z_1} =  \frac{u_*}{\kappa\,z_1} \phi_m(z_1/L).
\end{equation}
Here, $\phi_m(\zeta)=1/x(\zeta)$ for $\zeta < 0$ and $\phi_m(\zeta)=1+b\zeta$ for $\zeta\geq 0$.
Projecting the gradient onto the wind directions, we obtain $(\partial U/\partial z)|_{z_1}=(\partial M/\partial z)|_{z_1} U_1/M_1$ and $(\partial V/\partial z)|_{z_1}=(\partial M/\partial z)|_{z_1} V_1/M_1$, respectively.

\section{Test cases}\label{sec:results}
To verify the correct implementation of JAX-SCM, we run three well-known benchmark cases that cover the neutral, stable, and convective atmospheric stability regimes, in order of increasing complexity.
The first case is the turbulent neutral Ekman layer discussed by \citet{andren1994} and presented in sect.~\ref{sec:res_a94}.
The case is synthetic and forced only by constant geostrophic wind and Coriolis force, but has zero surface fluxes.
As a stable test case, we take the GABLS1 SCM intercomparison of \citet{cuxart2006} and compare JAX-SCM to their results in sect.~\ref{sec:res_gabls1}.
The case is a dry synthetic case with constant surface cooling leading to increasing stratification.
Finally, day 33 of the 1967 Wangara field campaign \citep{clarke1971} is reproduced and discussed in sect.~\ref{sec:res_wg33}.
This case is the most complex and also the most realistic of the three.
The simulation is initialized with observed wind, temperature, and humidity profiles and forced with time-varying surface sensible heat and moisture fluxes.
This case demonstrates the real-world applicability of JAX-SCM.

\subsection{\citet{andren1994}: neutral turbulent Ekman layer}\label{sec:res_a94}
\begin{figure*}[t]
    \includegraphics[width=\textwidth]{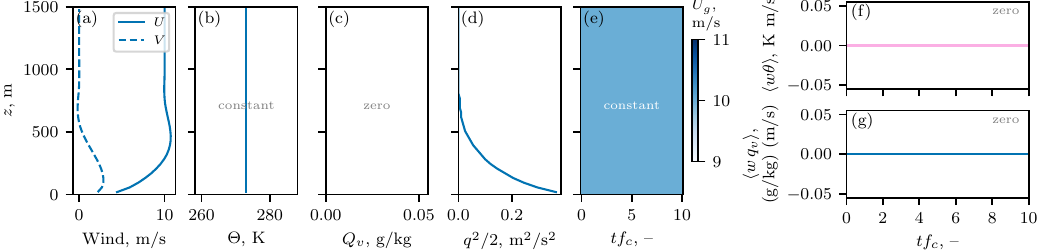}
    \caption{Initial and boundary conditions for the neutral turbulent Ekman layer test case \citep{andren1994}. Initial wind (a) and TKE (d) profiles are prescribed together with constant geostrophic forcing (e). All other initial profiles and surface forcings are zero (constant for $\Theta$). Domain height is $H=1500$\,m, discretized into $N=100$ levels ($\Delta z=15$\,m).}%
    \label{fig:ic_bc_a94}
\end{figure*}
\begin{figure}[t]
    \includegraphics[width=.9\columnwidth]{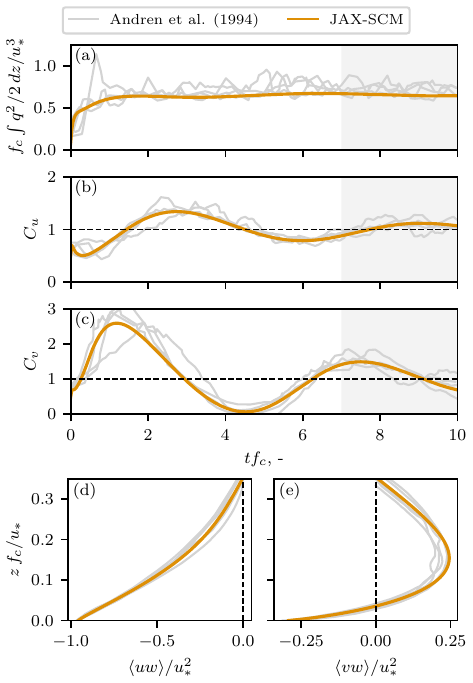}
    \caption{Comparison of JAX-SCM and \citet{andren1994} reference results in terms of (a) column-integrated normalized TKE, normalized deviation from steady state of mean wind profiles of (b) $U$ and (c) $V$, and (d+e) averaged normalized momentum flux profiles. Averaging window of momentum fluxes (d+e) is $t\,f_c \in [7, 10]$ as visualized by grey shading in (a)--(c).}%
    \label{fig:res_a94}
\end{figure}

The goal of the neutral turbulent Ekman layer is to evaluate how JAX-SCM behaves in a simple synthetic case: a flow driven by a constant large-scale pressure gradient in a rotating system \citep[A94 from now on]{andren1994}. 
The large-scale pressure gradient is imposed in the simulation as a constant geostrophic wind forcing of $U_g=10$\,m\,s$^{-1}$.
The Coriolis frequency is set to $f_c=10^{-4}$, corresponding to ca.~45$^{\circ}$N latitude, and the surface has a prescribed roughness of $z_{0m}=0.1$\,m.
As this case targets neutral atmospheric stratification, no sensible heat and moisture fluxes are prescribed at the surface.
The initial conditions for $u$, $v$, and $q^2$ profiles are taken from Table~A.1 of A94 and are visualized in Fig.~\ref{fig:ic_bc_a94}, panels (a)--(d).
The constant forcings are presented in panels (e), (f), and (g).
The domain height is set to $H=1500$\,m and is discretized into $N=100$ full levels, leading to $\Delta z=15$\,m.
The simulation is run for 10 inertial periods, i.e., $t=10/f_c \approx 27.78$\,h.

The simulated output of JAX-SCM (orange) is compared to the ensemble of simulations intercompared in A94 (grey) in Fig.~\ref{fig:res_a94}.
Only plots relevant to verifying the SCM performance are presented here, and discussions in A94 related to the original benchmark of large eddy simulation (LES) are skipped.
We also treat the four codes in A94 as a single reference ensemble, as we only aim to verify the correct implementation of JAX-SCM.

Panel (a) of Fig.~\ref{fig:res_a94} shows the column-integrated TKE, normalized with $f_c$ and the surface friction velocity $u_*$ from MOST;
panels (b) and (c) assess the convergence of the simulation toward a steady state solution by plotting the non-stationarity of the velocity fields at every time step as 
\begin{subequations}
    \begin{align}
        C_u &= -\frac{f_c}{\langle u w \rangle_s} \int_{0}^H \left(V-V_g\right) dz \\
        C_v &= +\frac{f_c}{\langle v w \rangle_s} \int_{0}^H \left(U-U_g\right) dz
    \end{align}
\end{subequations}
where $\langle u w \rangle_s$ and $\langle v w \rangle_s$ are the surface momentum fluxes from MOST;
and panels (d) and (e) present the normalized vertical profiles of momentum fluxes.
Following A94, the momentum flux profiles are averaged over the last three inertial periods as indicated by the grey shading in panels (a)--(c).
In all five panels, JAX-SCM agrees closely with the A94 reference simulation. 
The JAX-SCM curves are smoother because the contribution of turbulence to the flow is fully parameterized, whereas the LES runs partially resolve turbulence, leading to greater variability.
Along this line of thought, it is particularly encouraging for future turbulence research that the integrated TKE and the MYNN-2.5-parameterized momentum flux profiles match the LES results so well.
In conclusion, JAX-SCM simulates the neutral Ekman layer very well, verifying the applicability to shear-driven flows.

\subsection{GABLS1: stable boundary layer}\label{sec:res_gabls1}
\begin{figure*}[t]
    \includegraphics[width=\textwidth]{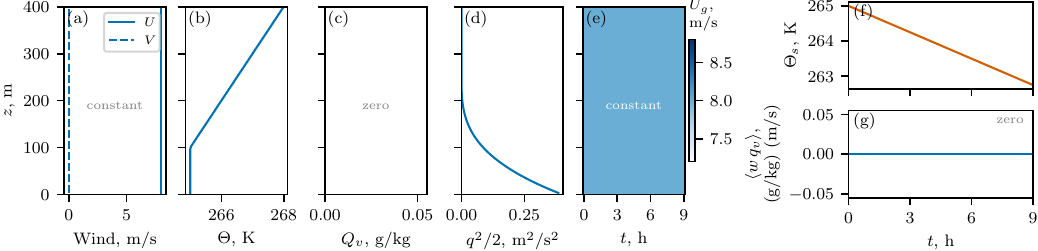}
    \caption{Initial and boundary conditions for dry stable boundary layer test case (GABLS1, \citet{cuxart2006}). Initial profiles for (a) wind, (b) potential temperature, and (d) TKE are prescribed together with (e) constant geostrophic wind and (f) constant cooling of the surface. Domain height is $H=400$\,m, discretized into $N=64$ levels ($\Delta z=6.25$\,m).}%
    \label{fig:ic_bc_gab1}
\end{figure*}
\begin{figure*}[t]
    \includegraphics[width=\textwidth]{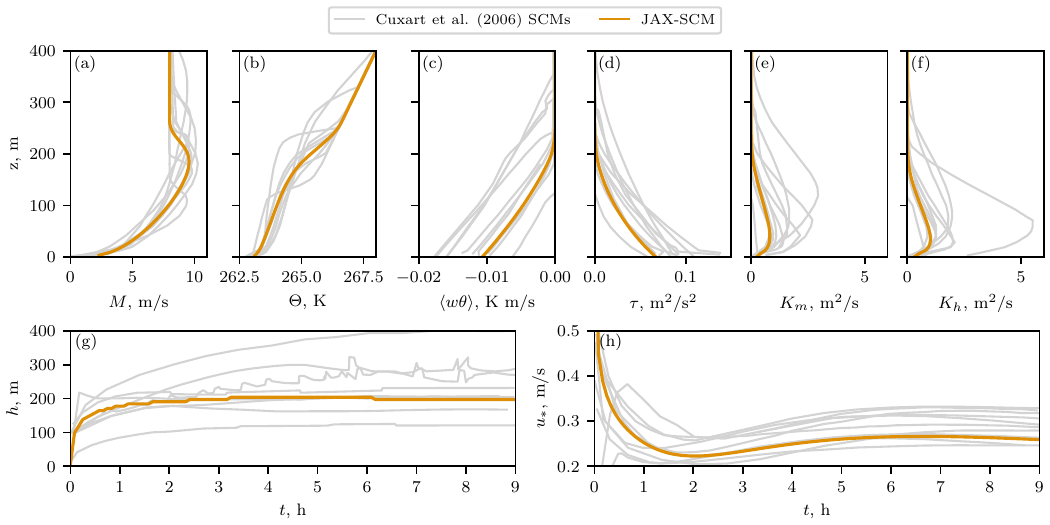}
    \caption{Comparision of JAX-SCM and GABLS1 SCM reference results \citep{cuxart2006} with respect to (a-f) profiles at $t=9h$ and temporal evolutions of (g) boundary layer height $h$ and (h) surface friction velocity. Panel (a) shows wind magnitude $M=\left(U^2+V^2\right)^{1/2}$ and panel (d) the total turbulent stress $\tau={\left({\langle u w\rangle}^2 + {\langle v w \rangle}^2\right)}^{1/2}$.}%
    \label{fig:res_gab1}
\end{figure*}

The next step is to assess JAX-SCM's performance in simulating a stably stratified atmospheric boundary layer.
The test case employed here comes from the GABLS1 SCM intercomparison conducted by \citet{cuxart2006}, who derived it from \citet{kosovic2000}.
Similar to A94, the simulation is subject to constant geostrophic forcing ($U_g=8$\,m\,s$^{-1}$) but experiences a constant cooling of $0.25$\,K\,h$^{-1}$ at the surface.
JAX-SCM directly supports this surface forcing as the user can prescribe either a heat flux or the surface temperature (used here) as the lower BC.
The initial profiles and forcings following \citet{cuxart2006} are visualized in Fig.~\ref{fig:ic_bc_gab1}.
The initial wind (panel a) is constant with height, humidity (panel c) is zero, and the potential temperature profile shows a neutral layer of depth 100\,m with stable stratification on top.
The simulation runs for 9\,h of simulated time in a 400\,m domain, discretized into $N=64$ levels.

The comparison of JAX-SCM against GABLS1 is presented in Fig.~\ref{fig:res_gab1}. 
As for A94, we do not consider the SCMs benchmarked in GABLS1 individually, but only verify JAX-SCM against the ensemble.
The profiles in panels (a)--(f) correspond to the state at the end of the simulation, after 9\,h, showing horizontal wind magnitude $M=\left(U^2+V^2\right)^{1/2}$, potential temperature, parameterized sensible heat flux, parameterized turbulent stress $\tau$, and the momentum and heat eddy diffusivities.
The two panels (g) and (h) show the temporal evolution of the boundary layer height $h$ and the surface friction velocity, as determined by MOST.
The total turbulent stress $\tau$ in panel (d) is computed from the parameterized vertical momentum flux profiles as
\begin{equation}
    \tau={\left({\langle u w\rangle}^2 + {\langle v w \rangle}^2\right)}^{1/2},
\end{equation}
and $h$ in panel (g) is defined as the height where $\tau$ vanishes.
More specifically, $h$ is computed by determining the height where $\tau$ falls to 5\% of its surface value, followed by a linear extrapolation to where $\tau \approx 0$ \citep{cuxart2006,beare2006}: $h = z_{\tau,5\%} / 0.95$ where $\tau(z)/\tau_s \approx 0.05$ is met at $z_{\tau,5\%}$. 

Similar to A94, the JAX-SCM outputs align well with the GABLS1 reference. 
All curves demonstrate the expected behavior, including the profiles of turbulent quantities ($\langle w\theta \rangle$, $\tau$, $K_m$, $K_h$) parameterized by MYNN-2.5.
Like the reference models, JAX-SCM successfully exerts surface friction on the initially constant wind profiles, decelerating wind at the surface and forming a jet at $z\approx180$\,m (cf.~Fig.~\ref{fig:res_gab1}a).
The constant surface cooling successfully yields a stratified atmosphere (cf.~Fig.~\ref{fig:res_gab1}b) and a realistic surface sensible heatflux of $\langle w\theta \rangle \approx-0.01$\,K\,m\,s$^{-1}$ (cf.~Fig.~\ref{fig:res_gab1}c).
JAX-SCM's option to prescribe the surface heat flux indirectly through the surface temperature is crucial for accurately simulating moderately stable to very stable conditions due to the duality of the heat flux \citep{malhi1995,basu2008}.

In conclusion, we consider the GABLS1 verification successful, demonstrating that JAX-SCM can simulate stable atmospheric boundary layers.

\subsection{Wangara, day 33: convective boundary layer}\label{sec:res_wg33}
\begin{figure*}[t]
    \includegraphics[width=\textwidth]{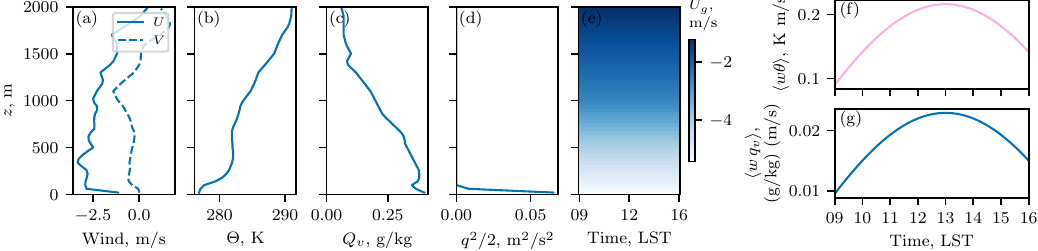}
    \caption{Initial and boundary conditions for convective boundary layer test case, day 33 of the Wangara field campaign \citep{clarke1971,nakanishi2009}. Initial profiles of (a) wind, (b) temperature, and (c) moisture are obtained from 09:00 soundings at Wangara. Simulation is forced with (e) constant-in-time geostrophic wind profiles and time-dependent (f) sensible heat flux and (g) moisture flux at the surface. Domain height is $H=2000$\,m, discretized into $N=100$ levels ($\Delta z = 20$\,m).}%
    \label{fig:ic_bc_wg33}
\end{figure*}
\begin{figure*}[t]
    \includegraphics[width=\textwidth]{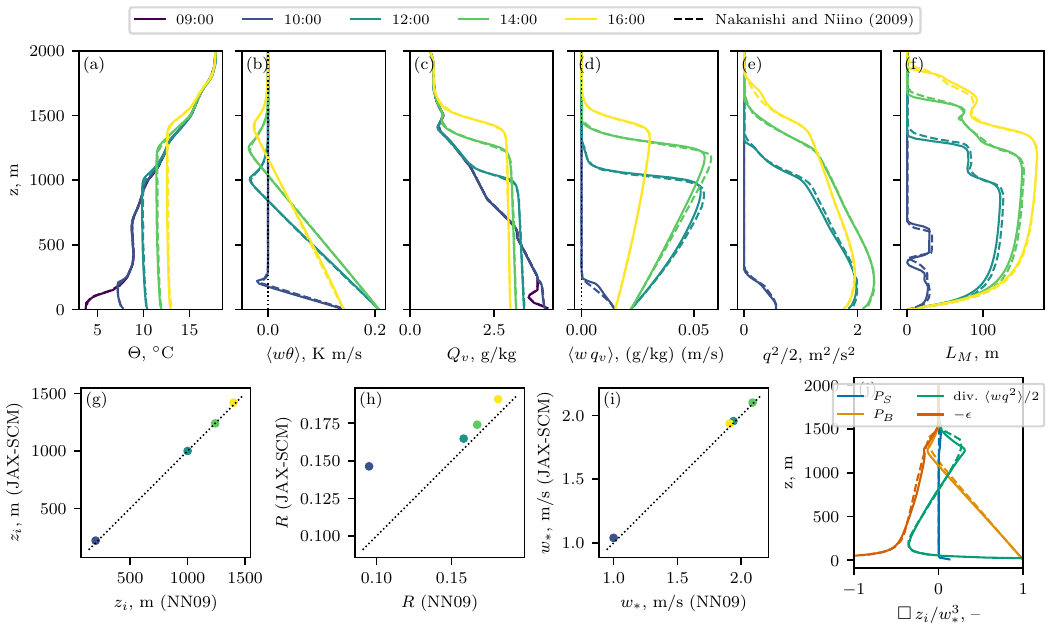}
    \caption{Comparision of JAX-SCM and reference results from \citet{nakanishi2009} for Wangara day 33. Temporal evolutions of profiles of mean and parameterized turbulent quantities are presented in (a) to (f). Mixed layer parameters are displayed in panels (g) to (i) with (g) inversion height $z_i$, (h) ratio $R$ of $\langle w \theta \rangle$ at $z_i$ and surface, (i) and convective velocity scale $w_*$. Panel (j) shows the normalized TKE budget over height at 14:00~LST.}%
    \label{fig:res_wg33}
\end{figure*}

The final test case is the simulation of a moist convective boundary layer with time-dependent surface forcing and observation-based initial conditions.
More specifically, we simulate day 33 of the 1967 Wangara field campaign \citep{clarke1971}, which has been used as a test case in numerous publications (e.g., \citet{wyngaard1974evolution,yamada1975,pielke1975representation,chen1983numerical,musson1995comparison,kim2003comparative,basu2008b}).
Crucially, Wangara Day 33 (WG33) is also the test case employed by \citet{nakanishi2009}, whose closure model JAX-SCM implements.

The initial conditions and forcings are summarized in Fig.~\ref{fig:ic_bc_wg33}.
Averaged soundings of mean wind, potential temperature, and specific humidity at 9:00 local time of campaign day 33 (16 August 1967) serve as initial conditions \citep[data page 227]{clarke1971}.
As neither \citet{clarke1971} nor NN09 specify an initial profile of $q^2$, we estimate a (twice) TKE surface value of $q^2$ as $q_s^2=B_1^{2/3} u_*^2$ \citep{mellor1982} and prescribe a vertical power-law decay of that value as
\begin{equation}
    q^2(z) = \begin{cases}
        q^2_s \left[1-\left(z/h_0\right)\right]^3 & z \leq h_0 \\
        0 & z > h_0.
    \end{cases}
\end{equation}
From the observed profile of $\Theta$ (cf.~Fig.~\ref{fig:ic_bc_wg33}b), we estimate an initial boundary layer height of $h_0=100$\,m for the equation above.
The TKE surface value is estimated as $q^2_s/2 \approx 0.1274$\,m$^2$\,s$^{-2}$ based on an observed $u_* \approx 0.175$\,m\,s$^{-1}$ \citep{hicks1981} and $B_1=24$ \citep{nakanishi2009}.
The domain height is set to $H=2000$\,m, in agreement with the maximum balloon height and previous studies.
We choose to discretize the domain into $N=100$ levels, leading to a grid spacing of $\Delta z=20$\,m.
Time integration is performed from 9:00 on day 33 until 16:00 on the same day, matching NN09.

Forcing enters the simulation through prescribed geostrophic wind, surface fluxes, and the Coriolis force.
Following NN09, the geostrophic wind forcing is height-dependent but constant in time (cf.~Fig.~\ref{fig:ic_bc_wg33}e), and the Coriolis parameter of $f_c \approx 8.26 \times 10^{-5}$\,s$^{-1}$ is set to match Wangara's latitude of 34.5$^{\circ}$S.
The surface fluxes of moisture and heat are time-dependent and take the form
\begin{equation}
    \langle w \phi \rangle_s = c \cos \left( \frac{t_h - 13}{11} \pi \right)
    \label{eq:wg33_sfc_forcing}
\end{equation}
where $t_h$ is the simulated time in hours since midnight of day 33, and $c=2.16 \times 10^{-1}$\,K\,m\,s$^{-1}$ for $\langle w\theta \rangle_s$ and $c=2.29 \times 10^{-5}$\,m\,s$^{-1}$ for $\langle w\, q_v\rangle_s$, respectively.

In contrast to the previous two cases, the WG33 test case is quite realistic.
The simulation is initialized from observations and subject to simplified but realistic time-dependent forcing.
Now, we also simulate the evolution of the specific humidity $Q_v$, verifying the final prognostic equation of JAX-SCM that has not been addressed so far.
Thus, this test employs all of JAX-SCM's features and also demonstrates its real-world applicability.

The comparison of JAX-SCM and the reference results from NN09 is presented in Fig.~\ref{fig:res_wg33}.
The first row displays the temporal evolution of profiles of potential temperature, specific humidity, their fluxes, TKE, and the master length scale throughout the day.
The reference solution of NN09 is given as dashed lines.
Panels (g)--(i) in the second row compare three mixed-layer parameters at different simulation times.
The inversion height $z_i$ in panel (g) is defined as the height where $\langle w \theta \rangle$ is minimal.
The strength of $\langle w \theta \rangle$ at that height relative to the surface value is quantified as $R=-\langle w \theta \rangle_{z_i}/\langle w \theta \rangle_{s}$ (cf.~panel (h)).
The convective velocity scale $w_*=(g/\Theta_0\, z_i \, \langle w \theta_v \rangle_s)^{1/3}$ with surface buoyancy flux $\langle w \theta_v \rangle_s$ is compared against NN09 values in (i).
Finally, panel (j) illustrates the TKE budget at 14:00, normalized by $z_i$ and $w_*$, respectively.

Overall, JAX-SCM matches all profiles very well, as expected given that it implements MYNN-2.5 according to NN09.
We attribute the minor differences visible between our code and the reference to the numerical implementation of the models.
It is also likely that our TKE initial profile differs from that used by NN09, leading to differences, especially at the beginning of the simulation (e.g., $\langle w\, q_v\rangle$ and $L_M$ at 10:00).
Similarly, the notable deviation of $R$ at 10:00 in Fig.~\ref{fig:res_wg33}h is attributed to model spin-up and differences in vertical discretization, which may affect the resolution of the thin low entrainment layer at the beginning of the simulation.
As the related values $z_i$ and $w_*$ are captured very well, we do not consider this deviation problematic.

In general, we view the WG33 test case as captured very well. 
Despite the minor differences discussed, the key behavior of the atmospheric column -- boundary layer growth, drying, and TKE budget -- is simulated realistically and accurately.
We consider the verification of JAX-SCM complete, having demonstrated the successful simulation of a neutral, a stably stratified, and a convective boundary layer.

\section{Technical details}\label{sec:tech_details}
\begin{figure*}[p]
\begin{lstlisting}[language=Python, caption=Setup of initial and boundary conditions for \citet{andren1994} neutral case., label=lst:a94_init]
import jax.numpy as jnp
import pandas as pd

from scm import convert
from scm.grid import StaggeredGrid
from scm.interfaces import Simulation, Forcing
from scm.mo import MOSettings
from scm.mynn.interfaces import ProgVarsMYNN

# Grid: 100 levels, 1500m domain height -> dz=15m
Nz = 100
grid = StaggeredGrid(Nz=Nz, H=1500)

# Compute Coriolis parameter at 45 degree latitude
f_c = convert.get_fc(lat_deg=45)  # ca. 1e-4 1/s

# Constant geostrophic wind
u_g = jnp.ones(Nz) * 10
v_g = jnp.zeros(Nz)

# Surface friction
mo_settings = MOSettings(z0h=0.1, z0m=0.1)

# Collect all forcing terms into a Forcing object
forcing = Forcing(
    u_geo=lambda t_s: u_g,
    v_geo=lambda t_s: v_g,
    f_c=f_c,
    w_th_s=lambda t_s: jnp.array(0.0),  # no surface heat flux
    w_qv_s=lambda t_s: jnp.array(0.0),  # no surface moisture flux
    dth_dz_top=0.0,  # no inversion at domain top
)

# Read initial profiles from csv file (Andren et al. 1994, Table A1)
df = pd.read_csv("ref/andren94_tab_A1.csv")

# Interpolate profiles to model grid
u = jnp.interp(grid.z, df["z"].values, df["u"].values)
v = jnp.interp(grid.z, df["z"].values, df["v"].values)
qke = jnp.interp(grid.z, df["z"].values, df["tke"].values) * 2  # QKE = 2 * TKE

# Collect all initial conditions into an object of prognostic variables
init = ProgVarsMYNN(
    u=u,
    v=v,
    th=jnp.ones(Nz) * 273.15,  # constant potential temperature profile
    qv=jnp.zeros(Nz),  # no humidity
    qke=qke,
)

# Assemble simulation object with all components
sim = Simulation(
    name="Andren1994",
    init=init,
    forcing=forcing,
    mo_settings=mo_settings,
    grid=grid,
    th_ref=273.15,
    t_start_s=0,  # simulation start time
    t_end_s=int(10 / f_c),  # simulate for 10 inertial periods
)
\end{lstlisting}
\end{figure*}

Having verified the correctness of JAX-SCM, we now discuss two technical aspects of its implementation that are relevant to users and developers.
The composable Python interface is presented in sect.~\ref{sec:tech_interface} using the A94 simulation as an example.
The choice of JAX as the numerical backend and its current and envisioned advantages are discussed in sect.~\ref{sec:tech_jax}.

\subsection{Modern composable Python interface}\label{sec:tech_interface}
\begin{figure}[t]
\end{figure}

The simple, composable Python interface of JAX-SCM is briefly presented by discussing the code needed to set up the A94 simulation (cf.~sect.~\ref{sec:res_a94}), which is given in listing~\ref{lst:a94_init}.

After importing the necessary packages, the script begins with setting up the staggered grid, followed by defining the forcings in lines~14--32.
As all forcings of the A94 case are constant in time, simple lambda functions are used to return the same vectors of geostrophic wind at every time step \texttt{t\_s} as well as constant values of zero for surface sensible heat flux and moisture flux.
For more complex forcings, such as the time-dependent sine-like forcing of WG33 (cf.~Eq.~(\ref{eq:wg33_sfc_forcing})), users can implement a simple Python function taking the current simulation time in seconds as input and returning the corresponding flux.
For the WG33 sensible heat flux, for example, the function could read
\begin{small}
\begin{verbatim}
def w_th_fn(t_s: float) -> float:
    t_h = t_s / 3600  # seconds to hours
    return 2.16e-1 * jnp.cos((t_h - 13) / 11 * jnp.pi)
\end{verbatim}
\end{small}

and would be assigned to \texttt{w\_th\_s} in l.~29 instead of the lambda expression.

The initial conditions are read from a text file using the pandas library (l.~35; \citet{pandas2026}) and interpolated to the model grid (ll.~37--40).
We stress that relying on a well-tested library like pandas simplifies our setup and enables JAX-SCM to ingest initial conditions in diverse formats.
The initial conditions, the forcings, and simulation metadata are all collected in one consistent \texttt{Simulation} object.
This object behaves like a tree, which can be passed around in Python code without the risk of ``losing'' or mismatching any settings along the way.

Once the simulation object is created, running JAX-SCM requires only a few steps:
\begin{lstlisting}[language=Python,basicstyle=\small]
from scm.config import load_namelist
from scm.io.local import out_to_ds
from scm.mynn.model import init_model
from scm.time_stepping import simulate

# Setup simulation and load config
sim = ...
cfg = load_namelist("namelist.yaml")

# Setup model and run simulation
model = init_model(sim, cfg) 
out = simulate(model=model, sim=sim, cfg=cfg) 

# Save output to netcdf
ds = out_to_ds(out, sim)  
ds.to_netcdf("andren1994.nc") 
\end{lstlisting}

After loading the namelist file (l.~8) and setting up the model (l.~11), the simulation is executed in line~12.
The namelist file enables the user, e.g., to configure the simulation output frequency or select the time integration method. 
For example, to configure the CN solver with an internal time step of $\Delta t = 0.5$\,s and an output frequency of 5\,min, \texttt{namelist.yml} looks like this: 
\begin{lstlisting}[basicstyle=\small]
time_int: "implicit"
dt_s: .5
dt_s_out: 300.0
\end{lstlisting}

The simulation outputs are converted to an xarray dataset in line~15.
Xarray \citep{hoyer2017} is another popular data-handling library that interfaces with various storage backends, such as netcdf (cf.~l.~16) or the modern zarr format \citep{zarr2026}.
Being able to use such an established library for data output is another benefit of the Python ecosystem.

Overall, we believe that JAX-SCM strikes a good balance between numerical performance via JAX and usability with Python.
The technical advantages of utilizing JAX are presented in the following section.

\subsection{GPU runs and automatic differentiation with JAX}\label{sec:tech_jax}
As motivated in the introduction, JAX-SCM is implemented in Python to improve accessibility and enable the use of modern GPU hardware and ML-ready software tools.
However, as an interpreted programming language, plain Python trades performance for ease of use, making it much slower than compiled languages like Fortran.
To still keep the numerical components fast and efficient, we leverage the computing library JAX \citep{jax2018github}, which performs just-in-time compilation of Python code into fast XLA (Accelerated Linear Algebra; \citet{xla}) code.
XLA supports compilation for multiple hardware backends, notably standard CPU and modern GPUs.
\Citet{hafner2021}, for example, document on-par CPU performance between JAX-based and Fortran-based versions of their ocean model and a ca.~2x speedup of the JAX-GPU version over Fortran.
Just-in-time compilation is transparent to the user, so JAX-SCM runs on CPUs and GPUs alike without any code changes.
While a single JAX-SCM run does not require the parallelization offered by GPUs, GPUs can greatly accelerate tasks such as parameter sweeps or ensemble simulations with perturbed initial conditions to quantify uncertainties.

Additionally, JAX natively supports automatic differentiation (AD), which enables computing derivatives of implemented functions with respect to any of their inputs, down to machine precision \citep{baydin2018}.
Currently, JAX-SCM is not yet fully differentiable, as special care must be taken to avoid breaking gradients \citep{meunier2025}.
In MYNN-2.5, for example, computing the turbulent velocity scale $q = \sqrt{q^2}$ breaks gradients near $q \to 0$, as $d/dx(\sqrt{x}) = 1/(2\sqrt{x})$ goes to infinity at $x=0$.
In contrast to standard machine learning functions, physical parameterizations are not designed with differentiability in mind and therefore require modifications to avoid such gradient issues.
We target full differentiability in future work, but would still like to briefly introduce AD as a key feature of JAX and motivate the envisioned use cases.

Gradients are computed by AD in two steps \citep{baydin2018}.
First, during model execution, the forward pass, the chain of elementary numerical operations (e.g., addition, division, exponentiation, sine, cosine, \ldots) used to produce the solution is stored in a computational graph.
Analytical derivatives of all these elementary operations are known and implemented in JAX.
To compute the gradient/derivatives of the solution with respect to the inputs, the computation graph is traced backward, and gradients are computed by evaluating the elementary derivatives and applying the chain rule.
This reverse-mode AD is equivalent to the adjoint state method, which is typically challenging to apply for models implemented in traditional languages.
In JAX, however, the adjoint is available ``for free'', provided all gradients are well-behaved.

Once JAX-SCM is fully differentiable, AD can be used, for example, to tune the SCM so that its simulated state trajectory more closely matches a reference trajectory \citep{gelbrecht2023}.
This is needed, for example, to tune the parameters of turbulence closure models efficiently or to perform data assimilation, where initial conditions are optimized to match observed states.
Moreover, such fully-differentiable numerical models enable the integration of machine learning components into the numerical solver (cf.~e.g.,~\citep{kochkov2021,kochkov2024a,sanderse2024}).
As the Python-JAX setup generally enables these applications, we believe that JAX-SCM will be a powerful platform for future hybrid physics--machine-learning approaches.

\section{Conclusion}\label{sec:conclusion}
We presented JAX-SCM v1.0, a modern open-source atmospheric single-column model (SCM) implemented in Python using the JAX library \citep{jax2018github}.
The model solves for mean wind, temperature, humidity, and turbulent kinetic energy (TKE) using the level-2.5 Mellor-Yamada-Nakanishi-Niino (MYNN-2.5) turbulence closure scheme \citep{nakanishi2009}, and its modular design allows alternative closure schemes to be integrated in the future.
The Python interface aims to be a simple, modern, and accessible interface for users and developers alike.
By choosing Python, JAX-SCM can be combined with the extensive ecosystem of scientific packages. 
The JAX backend adds performance competitive with compiled Fortran via just-in-time compilation \citep{hafner2021}, transparent GPU execution for computationally demanding tasks such as parameter sweeps and ensemble simulations, and native automatic differentiation (AD) as a path toward future hybrid physics--machine learning approaches \citep{gelbrecht2023}.

We verified the correct implementation of JAX-SCM by simulating three well-known benchmark cases: the neutral turbulent Ekman layer \citep{andren1994}, the dry stable boundary layer of GABLS1 \citep{kosovic2000,cuxart2006}, and the convective boundary layer of Wangara day 33 \citep{clarke1971,yamada1975}.
JAX-SCM matched the reference solutions closely in all three cases, confirming its correct implementation across all atmospheric stability regimes.
Parameterized turbulent quantities, such as heat and momentum fluxes and TKE, were well captured throughout, establishing JAX-SCM as an immediately applicable tool for atmospheric boundary layer research.
The close agreement with Wangara day 33, initialized from observed profiles and forced with realistic time-dependent surface fluxes, further demonstrates its real-world applicability.

Looking ahead, two main directions are planned for future releases.
First, a partial-condensation scheme (e.g., \citet{nakanishi2009}) will be implemented to enable cloud formation, opening JAX-SCM to the historically central use case of developing and testing cloud parameterizations \citep{betts1986,randall1996}.
Second, and most importantly, we aim to make JAX-SCM fully differentiable.
Once differentiable, JAX-SCM will support gradient-based parameter optimization and data assimilation, as well as direct integration of machine-learning components within the solver \citep{gelbrecht2023}.
We envision similar hybrid physics--ML approaches demonstrated at the global scale \citep{kochkov2024a,kochkov2021} also for the atmospheric boundary layer.

In conclusion, JAX-SCM is a modern and accessible SCM that is immediately applicable to boundary layer research and is designed from the ground up as a platform for future hybrid physics--machine-learning approaches in the atmospheric boundary layer.

\section*{Code and data availability}
JAX-SCM is available on GitHub (\url{https://github.com/mpierzyna/jax_scm}) or on Zenodo (\url{https://doi.org/10.5281/zenodo.20314409}).
The validation data were digitized without modification from the respective original publications and are also available on GitHub and Zenodo.

\section*{Author contribution}
MP has conceived the idea for JAX-SCM, implemented it, validated it, and wrote the manuscript.

\section*{Competing interests}
MP does not declare any competing interests.

\section*{Acknowledgements}
MP receives funding from the FREE project (P19-13) of the research programme TTW-Perspectief which is (partly) financed by the Dutch Research Council (NWO).
MP is grateful to Sukanta Basu, Yi Dai, and André van Ginkel for discussions and feedback regarding JAX-SCM.
Generative large language models were used to edit the draft manuscript and to optimize the JAX-SCM codebase.

\bibliographystyle{apsrev4-2}
\bibliography{references.bib}

\end{document}